\documentclass[aps, twocolumn, pra, showpacs, superscriptaddress]{revtex4}
\usepackage{graphicx}
\usepackage{dcolumn}
\usepackage{bm}

\begin{document}

\title{Construct order parameter from the spectra of mutual information}
\author{Shi-Jian Gu}
\altaffiliation{Email: sjgu@phy.cuhk.edu.hk}
\affiliation{Department of Physics and ITP, The Chinese University of Hong Kong, Hong
Kong, China}
\author{Wing Chi Yu}
\affiliation{Department of Physics and ITP, The Chinese University of Hong Kong, Hong
Kong, China}
\author{Hai-Qing Lin}
\affiliation{Department of Physics and ITP, The Chinese University of Hong Kong, Hong
Kong, China}
\affiliation{Beijing Computational Science Research Center, Beijing 100084, China}

\begin{abstract}
In this paper, we try to establish a connection between a quantum
information concept, i.e. the mutual information, and the conventional order
parameter in condensed matter physics. We show that a non-vanishing mutual
information at a long distance means the existence of long-range order. By
analyzing the entanglement spectra of the reduced density matrix that are
used to calculate the mutual information, we show how to find the local
order operator used to identify various phases with long-rang order.
\end{abstract}

\maketitle







\section{Introduction}

\bigskip

At absolute zero temperature, a quantum many-body system may undergo a phase
transition as the system's parameter varies across a critical point\cite%
{Sachdev}. Except for some unconventional phase transitions, continuous
quantum phase transitions can be well characterized in the framework of the
Landau's symmetry-breaking theory. In the Landau's theory, the concept of
the order parameter plays a central role. To characterize a given ordered
phase, people usually introduce an order parameter which is nonzero in the
symmetry-breaking phase, but vanishes in the other phases. In the recent
years, much attention has been paid to using quantum information \cite%
{Nielsen1} approaches to investigate quantum phase transitions. A typical
example is the role of entanglement in quantum critical phenomena \cite%
{AOsterloh2002,TJOsbornee,SJGuXXZ,SJGUPRL,Aanfossi05,SJGuJPA}. Since the
quantum phase transitions occur at the absolute zero temperature, they are
purely driven by quantum fluctuations. People believe that the quantum
entanglement, as a kind of quantum correlation, should play a very important
role in the quantum phase transitions. This point has been proven to be true
in the later hundreds of works on the issue(for a review, see \cite{LAmico08}
and the references therein). The entanglement manifests interesting
properties, such as scaling \cite{AOsterloh2002,TJOsbornee}, singularity, or
maximum\cite{SJGuXXZ,SJGUPRL} etc, in various quantum phase transitions.

However, it seems to us that, besides its interesting behavior around the
critical point, the entanglement actually cannot tell us much information
about the corresponding phase itself. That is, while the entanglement helps
us to witness the quantum phase transition, a non-trivial question is how
can we learn the corresponding order parameter of the ordered phases from
the behavior of the entanglement. This question leads to the main motivation
of the present work. We will show that the non-vanishing mutual information
\cite{Nielsen1} at a long distance not only means the existence of
long-range order, but also can tell us the potential order parameter if we
analyze the entanglement spectra of the corresponding reduced density
matrices. A possible scheme to construct the order parameter will be
discussed in detail.

On the other hand, the order parameter plays a very important role in
condensed matter physics. In order to study the ground-state properties of a
quantum many-body system, people usually assume a special order parameter in
various methods. For instance, in the dynamic mean-field theory, a suitable
local order parameter usually is assumed at the very beginning. However,
such an order parameter strongly depends on the physical intuition of the
investigator. That means, if the guessed order parameter is wrong, then all
analysis based on the assumption of the order parameter are not physical or
even wrong. Therefore, a scheme that can tell us how to find an appropriate
order parameter is instructive for many condensed matter theorists in their
studies.

The paper is organized as follows. In Sec. II, we firstly discuss the
relation between the existence of long-range order and the non-vanishing
behavior of the mutual information. In Sec. III, we propose a possible
scheme to find the local order operators by analyzing the entanglement
spectra of the mutual information. Three simple applications are given in
the same section. Finally, we give a discussion and a brief summary in
section IV.

\section{Non-vanishing mutual information and long-range order}

Let us consider a general quantum many-body system described by the
Hamiltonian $H=\sum_{i}h_{i},$where $h_{i}$ is the local Hamiltonian. $h_{i}$
can be the kinetic energy, interaction, etc. To learn the potential
long-range correlations existing in the system's ground state $\left\vert
\Psi \right\rangle $, we focus on two small blocks, $i$ and $j$, separated
by a distance $|i-j|$. Without loss of generality, we assume that the block
has a minimum size with which the block can capture the potential long-range
correlation existing in the system, and the distance $|i-j|$ is comparable
to the system's size. The reduced density matrix of the block $i$ can be
expressed as
\begin{equation}
\langle \mu ^{\prime }|\rho _{i}|\mu \rangle =\mathrm{tr}(a_{i\mu ^{\prime
}}\rho a_{i\mu }^{\dagger }),
\end{equation}%
where $\rho =\left\vert \Psi \right\rangle \left\langle \Psi \right\vert $,
tr is used to trace out all other degrees of freedom of the system except
for the block $i$, $a_{i\mu }$ is the annihilation operator for the state $%
|\mu \rangle $ localized at the block $i$. The operators $a_{i\mu }$s
satisfy the commutation (or anti-commutation) relation for bosonic (or
fermionic) states. Similarly we can define the reduced density matrix of the
two blocks $i$ and $j$ as%
\begin{equation}
\langle \mu ^{\prime }\nu ^{\prime }|\rho _{i\cup j}|\mu \nu \rangle =%
\mathrm{tr}(a_{i\mu ^{\prime }}a_{j\nu ^{\prime }}\rho a_{j\nu }^{\dagger
}a_{i\mu }^{\dagger }).
\end{equation}%
The reduced density matrices are positive semidefinite and can be normalized
as
\begin{equation}
\mathrm{tr}(\rho _{i})=\mathrm{tr}(\rho _{j})=\mathrm{tr}(\rho _{i\cup j})=1.
\end{equation}%
So if we diagonalize these matrices%
\begin{eqnarray}
\rho _{i} &=&\sum_{\mu }p_{\mu }|\varphi _{i\mu }\rangle \langle \varphi
_{i\mu }|, \\
\rho _{j} &=&\sum_{\nu }p_{\nu }|\varphi _{i\nu }\rangle _{{}}\langle
\varphi _{i\nu }|, \\
\rho _{i\cup j} &=&\sum_{\mu \nu }q_{\mu \nu }|\phi _{i\mu ,j\nu }\rangle
\langle \phi _{i\mu ,j\nu }|,
\end{eqnarray}%
for a given reduced density matrix, say $\rho _{i}$, the diagonal elements $%
\{p_{\mu }\}$ then have a probability interpretation in the corresponding
eigenstate space. The von-Neumann entropy%
\begin{equation}
S=-\sum_{\mu }p_{\mu }\log _{2}p_{\mu }
\end{equation}%
measures the entanglement between the block $i$ and the rest of the system.

To study the correlation between the block $i$ and $j$, we introduce the
mutual information%
\begin{equation}
S(i|j)=S\left( {\rho }_{i}\right) +S\left( {\rho }_{j}\right) -S\left( {\rho
}_{i\cup j}\right) .
\end{equation}%
In the quantum information science, the mutual information measures the
total correlation between the two blocks. It has also been used to study
quantum critical phenomena in recent years\cite{Aanfossi05,SJGuJPA}. For the
completeness of the work, we show that the non-vanishing behavior of the
mutual information at a long distance leads to the existence of long-range
order, as discussed in Ref. \cite{SJGuJPA}. For this purpose, we express the
mutual information as
\begin{equation}
S(i|j)=\mathrm{tr}(\rho _{i\cup j}\log _{2}\rho _{i\cup j})-\mathrm{tr}(\rho
_{i\cup j}\log _{2}\rho _{i}\otimes \rho _{j}).
\end{equation}%
Then%
\begin{eqnarray}
&&S(i|j)=\sum_{\mu \nu }q_{\mu \nu }\log _{2}q_{\mu \nu }  \nonumber \\
&&\;\;\;\;\;\;\;\;\;-\sum_{\mu \nu }\langle \phi _{i\mu ,j\nu }|\rho _{i\cup
j}\log _{2}(\rho _{i}\otimes \rho _{j})|\phi _{i\mu ,j\nu }\rangle ,
\end{eqnarray}%
in the basis of $|\phi _{\mu \nu }\rangle $. Insert the identity $\sum_{\mu
\nu }|\varphi _{\mu }\varphi _{\nu }\rangle \langle \varphi _{\nu }\varphi
_{\mu }|=1$, the mutual information becomes
\begin{equation}
S(i|j)=\sum_{\mu \nu }q_{\mu \nu }\log _{2}q_{\mu \nu }-\sum_{\mu \nu \mu
^{\prime }\nu ^{\prime }}P_{\mu ^{\prime }\nu ^{\prime },\mu \nu }q_{\mu
^{\prime }\nu ^{\prime }}\log _{2}(p_{\mu }p_{\nu })
\label{eq:relativeentropy1}
\end{equation}%
with
\begin{equation}
P_{\mu ^{\prime }\nu ^{\prime },\mu \nu }=\langle \phi _{i\mu ,j\nu
}|\varphi _{i\mu ^{\prime }}\varphi _{j\nu ^{\prime }}\rangle \langle
\varphi _{i\mu ^{\prime }}\varphi _{j\nu ^{\prime }}|\phi _{i\mu ,j\nu
}\rangle  \label{eq:projection}
\end{equation}%
which satisfies
\begin{equation}
P_{\mu ^{\prime }\nu ^{\prime },\mu \nu }>0,\;\;\sum_{\mu ^{\prime }\nu
^{\prime }}P_{\mu ^{\prime }\nu ^{\prime },\mu \nu }=1,\;\sum_{\mu \nu
}P_{\mu ^{\prime }\nu ^{\prime },\mu \nu }=1.
\end{equation}%
Eq. (\ref{eq:relativeentropy1}) can be rewritten as
\begin{equation}
S(i|j)=\sum_{\mu \nu }q_{\mu \nu }\left( \log _{2}q_{\mu \nu }-\sum_{\mu
^{\prime }\nu ^{\prime }}P_{\mu \nu ,\mu ^{\prime }\nu ^{\prime }}\log
_{2}(p_{\mu ^{\prime }}p_{\nu ^{\prime }})\right) .
\end{equation}%
Now we divide the problem into the following two cases.

Case I): For any matrix element in $\rho _{i\cup j}$, if we can always find $%
\langle a_{i\mu }^{\dagger }a_{i\mu ^{\prime }}a_{j\nu }^{\dagger }a_{j\nu
^{\prime }}\rangle =\langle a_{i\mu }^{\dagger }a_{i\mu ^{\prime }}\rangle
\langle a_{j\nu }^{\dagger }a_{j\nu ^{\prime }}\rangle $ such that $\rho
_{i\cup j}=\rho _{i}\otimes \rho _{j}$, then $P_{\mu ^{\prime }\nu ^{\prime
},\mu \nu }$ is a unit matrix and we have $S(i|j)=0$. In this case, any
linear superposition of $a_{i\mu }^{\dagger }a_{i\mu ^{\prime }}$ is not
correlated, so the system does not have any long-range order.

Case II): If there exist some elements that $\langle a_{i\mu }^{\dagger
}a_{i\mu ^{\prime }}a_{j\nu }^{\dagger }a_{j\nu ^{\prime }}\rangle \neq
\langle a_{i\mu }^{\dagger }a_{i\mu ^{\prime }}\rangle \langle a_{j\nu
}^{\dagger }a_{j\nu ^{\prime }}\rangle $, then $\rho _{i\cup j}\neq \rho
_{i}\otimes \rho _{j}$,\ $P_{\mu ^{\prime }\nu ^{\prime },\mu \nu }$ in Eq. (%
\ref{eq:projection}) is no longer a unit matrix. Since the log function is
concave,
\[
\sum_{\mu ^{\prime }\nu ^{\prime }}P_{\mu ^{\prime }\nu ^{\prime },\mu \nu
}\log _{2}(p_{\mu ^{\prime }}p_{\nu ^{\prime }})<\log _{2}\left( \sum_{\mu
^{\prime }\nu ^{\prime }}P_{\mu ^{\prime }\nu ^{\prime },\mu \nu }p_{\mu
}p_{\nu }\right) ,
\]%
we have
\begin{eqnarray}
S(i|j) &>&\sum_{\mu \nu }q_{\mu \nu }\log _{2}Q_{\mu \nu }  \nonumber \\
&>&\frac{1}{\ln 2}\sum_{\mu \nu }q_{\mu \nu }\left( 1-Q_{\mu \nu }\right) =0,
\end{eqnarray}%
wherewhere
\[
Q_{\mu \nu }=\frac{q_{\mu \nu }}{\sum_{\mu ^{\prime }\nu ^{\prime }}P_{\mu
^{\prime }\nu ^{\prime },\mu \nu }p_{\mu }p_{\nu }}.
\]%
So the mutual information is non-vanishing
\begin{equation}
S(i|j)>0.
\end{equation}%
In this case, $\rho _{i\cup j}\neq \rho _{i}\otimes \rho _{j}$. Actually, $%
S(i|j)>0$ is the sufficient and necessary condition for the existence of the
correlation between two blocks. That is, if and only if $S(i|j)>0$, we can
find matrix elements satisfying%
\begin{equation}
\langle a_{i\mu }^{\dagger }a_{i\mu ^{\prime }}a_{j\nu }^{\dagger }a_{j\nu
^{\prime }}\rangle \neq \langle a_{i\mu }^{\dagger }a_{i\mu ^{\prime
}}\rangle \langle a_{j\nu }^{\dagger }a_{j\nu ^{\prime }}\rangle .
\end{equation}%
Then $\langle a_{i\mu }^{\dagger }a_{i\mu ^{\prime }}\rangle $ plays a role
of local order operator at site $i$. Nevertheless, there are usually many
terms satisfying the above inequality, the operator $\langle a_{i\mu
}^{\dagger }a_{i\mu ^{\prime }}\rangle $ is not a good candidate for
long-range order parameter.

To find a good candidate of the order parameter, we need to analyze spectra
of the mutual information. To proceed, we would like to mention some
properties of the mutual information:

i) If the rank of ${\rho }_{i}$ is 1, then there is no correlation between ${%
\rho }_{i}$ and ${\rho }_{j}$.

Prove: If the rank of ${\rho }_{i}$ is 1, then ${\rho }_{i}$ is a pure state
and $S({\rho }_{i})=0$. Since $S(i|j)=S\left( {\rho }_{i}\right) +S\left( {%
\rho }_{j}\right) -S\left( {\rho }_{i\cup j}\right) $ is non-negative, so $%
S(i|j)=0$. There is no correlation between the blocks $i$ and $j$.

ii) If the rank of ${\rho }_{i}$ is $\xi $, for any\ $\mu >\xi $, the
operator $a_{i\mu }^{\dagger }a_{i\mu }$ does not correlated.

Prove: If%
\begin{equation}
{\rho }_{i}|\varphi _{i\mu }\rangle =0,
\end{equation}%
Then%
\begin{equation}
\langle a_{i\mu }^{\dagger }a_{i\mu }\rangle =0.
\end{equation}%
The correlation function becomes%
\begin{eqnarray}
&&\langle a_{i\mu }^{\dagger }a_{i\mu }a_{j\mu }^{\dagger }a_{j\mu }\rangle
\nonumber \\
&=&\sum_{\mu ^{\prime }\nu ^{\prime }}\langle a_{i\mu }^{\dagger }a_{i\mu
}|\varphi _{i\mu ^{\prime }}\varphi _{j\nu ^{\prime }}\rangle \langle
\varphi _{i\mu ^{\prime }}\varphi _{j\nu ^{\prime }}|a_{j\mu }^{\dagger
}a_{j\mu }\rangle  \nonumber \\
&=&0.
\end{eqnarray}

iii) For $\mu \neq \nu $, $\langle a_{i\mu }^{\dagger }a_{i\nu }\rangle
=\langle a_{i\nu }^{\dagger }a_{i\mu }\rangle =0$. If $\left\langle a_{i\mu
}^{\dagger }a_{i\nu }a_{j\mu }^{\dagger }a_{j\nu }\right\rangle \neq 0$, $%
a_{i\mu }^{\dagger }a_{i\nu }+a_{i\nu }^{\dagger }a_{i\mu }$ is a good
candidate of the order parameter.

These three properties will be used in the next section to construct local
order operator.

\section{Construct local order operator}

In this section, we show how to construct the potential local order
parameter from the spectra of the reduced density matrices. We have show
that the long-range order exists if and only if the mutual information does
not vanish at long distance. However, the mutual information can not tell us
directly which kind of order exists in the phase. To find the potential
order parameter, we need to look into the spectra of the mutual information.

At the beginning, we need to decide the minimum size of the two correlated
blocks. If the block size is too large, even if the mutual information does
not vanish, the final order parameter obtained might be too complicated to
be well understood. On the other hand, if the block size is too small, the
corresponding mutual information might not be able to find the order
parameter. For instance, in the spin dimer or trimer ordered phase, the
mutual information between two single spins usually decays exponentially
because the spin-spin correlation function decays exponentially too.
Therefore, to decide an appropriate block size is important. For this
purpose, we can start from a relatively large block size. In this case, it
is easier to witness if there exist long-range order in the system. Then we
reduce the size of the blocks step by step until find the minimum size of
the correlated blocks. Once the minimum block size is decided, we continue
to find the potential order parameters from the mutual information of the
two blocks.

\textit{Diagonal long-range order:} Usually, there are two kinds of
long-range order. One is the diagonal long-range order, the another is the
off-diagonal long-range order. Whether the order is diagonal or off-diagonal
depends on the basis used to define the reduced density matrices. Without
loss of generality, we use the eigenstates $|\varphi _{i\mu }\rangle $ of ${%
\rho }_{i}$ to define the local modes. Then the set $\{p_{i\mu }\}$ is the
probability distribution of these modes.

We now consider the first case of diagonal long-range order. In this case,
if we express the combined block $\rho _{i\cup j}$ in the basis $|\varphi
_{i\mu }\varphi _{j\nu }\rangle $, we can find that there exist some
diagonal elements $q_{\mu \nu }\neq p_{i\mu }p_{i\upsilon }$. We have shown
that if the rank of ${\rho }_{i}$ is $\xi $, for any\ $\mu >\xi $, the
operator $a_{i\mu }^{\dagger }a_{i\mu }$ does not correlated. Therefore, the
potential order operator can be constructed as a superposition of nonzero
diagonal elements of ${\rho }_{i}$
\begin{equation}
O_{i}=\sum_{\mu \leq \xi }w_{\mu }a_{i\mu }^{\dagger }a_{i\mu }.
\end{equation}%
Clearly, the definition still makes the order parameter not unique
especially if the rank of ${\rho }_{i}$ is large. To have an explicit form,
we can apply various conditions on the set of coefficient $\{w_{\mu }\}$.

The first condition can be the traceless condition, i.e.,
\begin{equation}
\sum_{\mu \leq \xi }w_{\mu }p_{\mu }=0,
\end{equation}%
which makes
\begin{equation}
\text{tr}\left( O_{i}\right) =0.
\end{equation}%
A special case is that the rank of ${\rho }_{i}$ is 1, then the traceless
condition will not leads to any order parameter. With the above condition,
we need only to calculate the function $\left\langle O_{i}O_{j}\right\rangle
$ for the correlation function.

The second condition can be the normalization condition. That is the maximum
value in the set $\{w_{\mu }\}$ is 1. This condition defines a maximum
amplitude of the correlation function $\left\langle O_{i}O_{j}\right\rangle $%
.

Usually, the above two conditions are enough to decide the form the order
parameter. In case that there are still some degrees of freedom in $\{w_{\mu
}\}$, one may fix the residue degrees of freedom by considering the other
related physical properties of the ground state.

Once the order parameter is obtained, the correlation function becomes%
\begin{eqnarray}
\left\langle O_{i}O_{j}\right\rangle &=&\sum_{\mu ^{\prime }\nu ^{\prime
}}\langle O_{i}|\varphi _{i\mu ^{\prime }}\varphi _{j\nu ^{\prime }}\rangle
\langle \varphi _{i\mu ^{\prime }}\varphi _{j\nu ^{\prime }}|O_{j}\rangle
\nonumber \\
&=&\sum_{\mu ,\nu \leq \xi }w_{\mu }w_{\nu }\sum_{\mu ^{\prime }\nu ^{\prime
}}p_{\mu ^{\prime }}p_{\nu ^{\prime }}P_{\mu ^{\prime }\nu ^{\prime },\mu
\nu }  \nonumber \\
&=&\sum_{\mu ,\nu \leq \xi ;\mu ^{\prime }\nu ^{\prime }}w_{\mu }w_{\nu
}p_{\mu ^{\prime }}p_{\nu ^{\prime }}P_{\mu ^{\prime }\nu ^{\prime },\mu \nu
}.
\end{eqnarray}%
Clearly, if $P_{\mu ^{\prime }\nu ^{\prime },\mu \nu }$ is a unity matrix, $%
\rho _{i\cup j}=\rho _{i}\otimes \rho _{j}$, $\left\langle
O_{i}O_{j}\right\rangle =0$, so there is no correlation. Otherwise,
\begin{equation}
\left\langle O_{i}O_{j}\right\rangle -\left\langle O_{i}\rangle \langle
O_{j}\right\rangle \neq 0,
\end{equation}%
then $O_{i}$ plays a role of order parameter.

\textit{Off-diagonal long-range order:} the off-diagonal long-range order
exists if the combined block $\rho _{i\cup j}$ is not diagonal in the basis $%
|\varphi _{i\mu }\varphi _{j\nu }\rangle $. For the block $i,$ since we
already have $\langle a_{i\mu }^{\dagger }a_{i\nu }\rangle =\langle a_{i\nu
}^{\dagger }a_{i\mu }\rangle =0$ for $\mu \neq \nu $, the nonzero
off-diagonal elements in $\rho _{i\cup j}$ means the existence of the
off-diagonal long-range order. To find the order parameter, the first step
is to find all possible indices pair $\left\langle \mu ,\nu \right\rangle
(\mu \neq \nu )$ with which $\left\langle a_{i\mu }^{\dagger }a_{i\nu
}a_{j\mu }^{\dagger }a_{j\nu }\right\rangle \neq 0$ and $\left\langle
a_{i\nu }^{\dagger }a_{i\mu }a_{j\mu }^{\dagger }a_{j\nu }\right\rangle \neq
0$. Then the order parameter can be expressed as%
\begin{equation}
O_{i}=\sum_{\left\langle \mu ,\nu \right\rangle }\left( w_{\mu \nu }a_{i\mu
}^{\dagger }a_{i\nu }+w_{\mu \nu }^{\ast }a_{i\nu }^{\dagger }a_{i\mu
}\right) .
\end{equation}%
Here we suppose the order parameter is a Hermition operator. From the
condition, we have tr$\left( O_{i}\right) =0$. We have noticed $\left\langle
a_{i\mu }^{\dagger }a_{i\nu }a_{j\mu }^{\dagger }a_{j\nu }\right\rangle \neq
0$ and $\left\langle a_{i\nu }^{\dagger }a_{i\mu }a_{j\mu }^{\dagger
}a_{j\nu }\right\rangle \neq 0$, so the correlation function $\left\langle
O_{i}O_{j}\right\rangle $ is nonzero too.

\textit{The mode of the order parameter:} Having obtained the exact form of
the order parameter, the next step is to determine the mode of the
corresponding correlation function. For this purpose, we reexpress $%
\left\langle O_{i}O_{j}\right\rangle $ as a function of $r=i-j$. Usually,
the correlation function is an oscillating function. If the wavelength of
the correlation function is $\lambda $. the mode of the order parameter is
\begin{equation}
k=\frac{2\pi }{\lambda }.
\end{equation}

Finally, we summarize our scheme in the following:

i) Via studying the mutual information between two blocks for various block
size, find the minimum block size with which the mutual information does not
vanish at a long distance.

ii) Calcuate the spectra $\{p_{\mu }\}$ of the reduced denisity matrix ${%
\rho }_{i}$ of the block $i$.

iii) Define the potential diagonal order parameter as%
\begin{equation}
O_{i}=\sum_{\mu \leq \xi }w_{\mu }a_{i\mu }^{\dagger }a_{i\mu },
\end{equation}%
where $\xi $ is the rank of the reduced density matrix, and $a_{i\mu
}^{\dagger }\left\vert 0\right\rangle $ defines the eigenvector of the
matrix with the eigenvalue $p_{\mu }$. Apply the traceless condition,
normalization condition, and other suitable restrictions, to determine the
set $\{w_{\mu }\}$.

iv) If the reduced density matrix of the combined block $\rho _{i\cup j}$ is
not diagonal in the basis of $\rho _{i}\otimes \rho _{j}$, define the
off-diagonal order parameter
\begin{equation}
O_{i}=\sum_{\left\langle \mu ,\nu \right\rangle }\left( w_{\mu \nu }a_{i\mu
}^{\dagger }a_{i\nu }+w_{\mu \nu }^{\ast }a_{i\nu }^{\dagger }a_{i\mu
}\right) ,
\end{equation}%
where the pairs $\left\langle \mu ,\nu \right\rangle $ can be determined
from nonzero off-diagonal elements in $\rho _{i\cup j}$. Determine the set $%
\{w_{\mu }\}$ with suitable conditions.

v) Calculate the correlation function $\left\langle O_{i}O_{j}\right\rangle $
as a function of $i-j$ to determine the mode of the order parameter.

\subsection{Application I: Ferromagnetic long-range order}

As a simple application, we take a spin chain with ferromagnetic long-range
order as an example. We consider such a state,%
\[
\left\vert \Psi \right\rangle =\frac{1}{\sqrt{2}}\left( \left\vert \uparrow
\uparrow \cdots \uparrow \uparrow \right\rangle +\left\vert \downarrow
\downarrow \cdots \downarrow \downarrow \right\rangle \right) .
\]%
For this state, we can find the minimum block size with which the mutual
information does not vanish is 1. The reduced density matrix of a single
spin is%
\begin{equation}
{\rho }_{i}=\left(
\begin{array}{cc}
1/2 & 0 \\
0 & 1/2%
\end{array}%
\right)
\end{equation}%
in the basis of $\{\left\vert \uparrow \right\rangle ,\left\vert \downarrow
\right\rangle \}$. For any two spins at site $i$ and $j$, the reduced
density matrix ${\rho }_{i\cup j}$
\begin{equation}
{\rho }_{i\cup j}=\left(
\begin{array}{llll}
1/2 & 0 & 0 & 0 \\
0 & 0 & 0 & 0 \\
0 & 0 & 0 & 0 \\
0 & 0 & 0 & 1/2%
\end{array}%
\right) ,
\end{equation}%
in the basis of $\{\left\vert \uparrow \uparrow \right\rangle ,\left\vert
\uparrow \downarrow \right\rangle ,\left\vert \downarrow \uparrow
\right\rangle ,\left\vert \downarrow \downarrow \right\rangle \}$.
Therefore,
\begin{equation}
S\left( {\rho }_{i}\right) =1,
\end{equation}%
and
\begin{equation}
S\left( {\rho }_{i\cup j}\right) =1.
\end{equation}%
The mutual information between any two spins becomes%
\begin{eqnarray}
S(i|j) &=&S\left( {\rho }_{i}\right) +S\left( {\rho }_{j}\right) -S\left( {%
\rho }_{i\cup j}\right) \\
&=&1.
\end{eqnarray}%
Therefore, there must exist long-range order in the state.

For the present state, ${\rho }_{i}$ is already diagonal and its spectra are
$\{1/2,1/2\}$. So we can define the order parameter as
\begin{equation}
O_{i}=w_{1}a_{i1}^{\dagger }a_{i1}+w_{2}a_{i2}^{\dagger }a_{i2}.
\end{equation}%
Then under the traceless condition%
\begin{equation}
\frac{1}{2}w_{1}+\frac{1}{2}w_{2}=0,
\end{equation}%
we have
\begin{equation}
w_{1}=-w_{2}.
\end{equation}%
Let $w_{1}=1$, we find a traceless operator for the local site%
\begin{equation}
O_{i}=a_{i1}^{\dagger }a_{i1}-a_{i2}^{\dagger }a_{i2},
\end{equation}%
which actually is the Jordan--Schwinger representation of the $z$-component
of Pauli matrix $\sigma _{i}^{z}$.

Having found the order parameter, the next step is to determine the mode of
the corresponding correlation function. From the two-site reduced density
matrix ${\rho }_{i\cup j}$, we have
\begin{eqnarray}
\langle a_{j1}^{\dagger }a_{i1}^{\dagger }a_{i1}a_{j1}\rangle &=&\frac{1}{2},
\\
\langle a_{j2}^{\dagger }a_{i2}^{\dagger }a_{i2}a_{j2}\rangle &=&\frac{1}{2},
\end{eqnarray}%
which is independent of the distance between $i$ and $j$. Therefore, the
correlation function is
\begin{equation}
\langle \sigma _{i}^{z}\sigma _{j}^{z}\rangle -\langle \sigma
_{i}^{z}\rangle \langle \sigma _{j}^{z}\rangle =1.
\end{equation}%
So in this case, the order parameter is $\sigma _{{}}^{z}$ with mode 0.

For a comparision, we consider another state with anti-ferromagnetic
long-range order,%
\[
\left\vert \Psi \right\rangle =\frac{1}{\sqrt{2}}\left( \left\vert \uparrow
\downarrow \cdots \uparrow \downarrow \right\rangle +\left\vert \downarrow
\uparrow \cdots \downarrow \uparrow \right\rangle \right) .
\]%
In a similar way, we can find that the order parameter is $\sigma _{{}}^{z}$%
. Neverthelesse, we notice that for even $|i-j|$
\begin{equation}
{\rho }_{i\cup j}=\left(
\begin{array}{llll}
1/2 & 0 & 0 & 0 \\
0 & 0 & 0 & 0 \\
0 & 0 & 0 & 0 \\
0 & 0 & 0 & 1/2%
\end{array}%
\right) ,
\end{equation}%
while for odd $|i-j|$
\begin{equation}
{\rho }_{i\cup j}=\left(
\begin{array}{llll}
0 & 0 & 0 & 0 \\
0 & 1/2 & 0 & 0 \\
0 & 0 & 1/2 & 0 \\
0 & 0 & 0 & 0%
\end{array}%
\right) .
\end{equation}%
So the correlation function becomes
\begin{eqnarray}
\langle \sigma _{i}^{z}\sigma _{j}^{z}\rangle &=&1,\text{ for even }|i-j| \\
\langle \sigma _{i}^{z}\sigma _{j}^{z}\rangle &=&-1,\text{for odd }|i-j|.
\end{eqnarray}%
In this case, the order parameter is still $\sigma _{{}}^{z}$, however, the
mode becomes $\pi $.

\subsection{Application II: Dimer order}

Now let us consider a uniformly weighted superposition of the two
nearest-neighbor valence bond state,
\begin{eqnarray}
&&|\psi _{1}\rangle =[1,2][3,4]\cdots \lbrack L-1,L]  \nonumber \\
&&|\psi _{2}\rangle =[L,1][2,3]\cdots \lbrack L-2,L-1]
\end{eqnarray}%
where
\[
\lbrack i,j]=\frac{1}{\sqrt{2}}(|\uparrow \rangle _{i}|\downarrow \rangle
_{j}-|\downarrow \rangle _{i}|\uparrow \rangle _{j}).
\]%
The inner product $\langle \psi _{2}|\psi _{1}\rangle $ decays expontially
with the system size, so the state we consider can be written as%
\begin{equation}
\left\vert \Psi \right\rangle =\frac{1}{\sqrt{2}}\left( |\psi _{1}\rangle
+|\psi _{2}\rangle \right) .
\end{equation}%
The state is the ground state of the Majumdar-Ghosh model\cite{CKMajumdar69}%
. For a single spin, the reduced density matrix is
\begin{equation}
{\rho }_{i}=\left(
\begin{array}{cc}
1/2 & 0 \\
0 & 1/2%
\end{array}%
\right)
\end{equation}%
For two spins separated by a long distance%
\begin{equation}
{\rho }_{i\cup j}=\left(
\begin{array}{llll}
1/4 & 0 & 0 & 0 \\
0 & 1/4 & 0 & 0 \\
0 & 0 & 1/4 & 0 \\
0 & 0 & 0 & 1/4%
\end{array}%
\right) .
\end{equation}%
Therefore $\rho _{i\cup j}=\rho _{i}\otimes \rho _{j}$, the $S(i|j)=0$.
There is no long-range spin-spin correlation. The results mean that the
block including only one spin is too small to find the long-range
correlation. So we let the block include two neighboring spins. The reduced
density matrix of two spins at site $i$ and $i+1$ is
\begin{equation}
{\rho }_{i}=\left(
\begin{array}{llll}
1/8 & 0 & 0 & 0 \\
0 & 3/8 & -1/4 & 0 \\
0 & -1/4 & 3/8 & 0 \\
0 & 0 & 0 & 1/8%
\end{array}%
\right)
\end{equation}%
The matrix can be diagonalized as%
\begin{eqnarray}
{\rho }_{i} &=&\frac{5}{8}\left\vert \varphi _{i,0}\right\rangle
\left\langle \varphi _{i,0}\right\vert +\frac{1}{8}\left\vert \varphi
_{i,1}\right\rangle \left\langle \varphi _{i,1}\right\vert  \nonumber \\
&&+\frac{1}{8}\left\vert \varphi _{i,2}\right\rangle \left\langle \varphi
_{i,2}\right\vert +\frac{1}{8}\left\vert \varphi _{i,3}\right\rangle
\left\langle \varphi _{i,3}\right\vert
\end{eqnarray}%
where%
\begin{eqnarray}
\left\vert \varphi _{i,0}\right\rangle &=&\frac{1}{\sqrt{2}}\left(
\left\vert \uparrow _{i}\downarrow _{i+1}\right\rangle -\left\vert
\downarrow _{i}\uparrow _{i+1}\right\rangle \right) , \\
\left\vert \varphi _{i,1}\right\rangle &=&\left\vert \uparrow _{i}\uparrow
_{i+1}\right\rangle , \\
\left\vert \varphi _{i,2}\right\rangle &=&\frac{1}{\sqrt{2}}\left(
\left\vert \uparrow _{i}\downarrow _{i+1}\right\rangle +\left\vert
\downarrow _{i}\uparrow _{i+1}\right\rangle \right) \\
\left\vert \varphi _{i,3}\right\rangle &=&\left\vert \downarrow
_{i}\downarrow _{i+1}\right\rangle .
\end{eqnarray}%
are spin singlet state and spin triplet states respectively. Then the
entropy
\[
S\left( {\rho }_{i}\right) =3-\frac{5}{8}\log _{2}5.
\]%
For any two pairs of spin, if the distance $|i-j|$ between them is even, the
density matrix is of dim $16\times 16$. In the basis of spin singlet and
spin triplet states, the reduced density matrix ${\rho }_{i\cup j}$ is
diagonal%
\begin{equation}
{\rho }_{i\cup j}=\frac{1}{2}\left\vert \varphi _{i,0}\varphi
_{j,0}\right\rangle \left\langle \varphi _{i,0}\varphi _{j,0}\right\vert +%
\frac{1}{32}I.
\end{equation}%
The entropy is
\[
S({\rho }_{i\cup j}{)}=5-\frac{17}{32}\log _{2}17.
\]%
The mutual information between the two blocks can be evaluated as
\[
S({i|j)}\simeq 0.269.
\]%
The nonzero value of the mutual information between two spin pairs means
that we can find the potential order parameter in their spectra. For this
purpose, we define it as%
\begin{eqnarray}
{O}_{i} &=&w_{0}\left\vert \varphi _{i,0}\right\rangle \left\langle \varphi
_{i,0}\right\vert +w_{1}\left\vert \varphi _{i,1}\right\rangle \left\langle
\varphi _{i,1}\right\vert  \nonumber \\
&&+w_{2}\left\vert \varphi _{i,2}\right\rangle \left\langle \varphi
_{i,2}\right\vert +w_{3}\left\vert \varphi _{i,3}\right\rangle \left\langle
\varphi _{i,3}\right\vert .
\end{eqnarray}%
Due to the symmetry among the three states $\{\left\vert \varphi
_{1}\right\rangle ,\left\vert \varphi _{2}\right\rangle ,\left\vert \varphi
_{3}\right\rangle \}$, we let $w_{1}=w_{2}=w_{3}$. Then the traceless
condition becomes%
\begin{equation}
\frac{5}{8}w_{0}+\frac{3}{8}w_{1}=0.
\end{equation}%
Upon the normalization condition, we let $w_{0}=-1$, we have%
\begin{eqnarray}
{O}_{i} &=&-\left\vert \varphi _{i,0}\right\rangle \left\langle \varphi
_{i,0}\right\vert +\frac{5}{3}\left\vert \varphi _{i,1}\right\rangle
\left\langle \varphi _{i,1}\right\vert  \nonumber \\
&&+\frac{5}{3}\left\vert \varphi _{i,2}\right\rangle \left\langle \varphi
_{i,2}\right\vert +\frac{5}{3}\left\vert \varphi _{i,3}\right\rangle
\left\langle \varphi _{i,3}\right\vert
\end{eqnarray}%
In terms of spin operators, the order parameter can be expressed as%
\begin{equation}
{O}_{i}=1+\frac{2}{3}\sigma _{i}\cdot \sigma _{i+1}.
\label{eq:spindimerorder}
\end{equation}%
The correlation function then can be calculated as%
\begin{eqnarray}
\left\langle {O}_{i}{O}_{j}\right\rangle &=&\frac{5}{8},\text{ for even }%
|i-j| \\
\left\langle {O}_{i}{O}_{j}\right\rangle &=&\frac{1}{2},\text{ for odd }%
|i-j|.
\end{eqnarray}%
The mode of the order parameter is therefore $\pi $.

Traditionally, people use the operator $\sigma _{i}\cdot \sigma _{i+1}$\cite%
{Singh} to describe the spin dimer order. Except for a difference in a
constant, the order parameter Eq. (\ref{eq:spindimerorder}) obtained via our
scheme is exactly the same as the traditional order parameter.

\subsection{Application III: Diagonal versus off-diagonal long-range
correlation}

To see the off-diagonal long-range correlation, we take the ground state of
the one-dimensional antiferromagnetic Heisenberg model as an example. The
model Hamiltonian reads%
\begin{equation}
H=\sum_{j=1}^{N}\left(
S_{j}^{x}S_{j+1}^{x}+S_{j}^{y}S_{j+1}^{y}+S_{j}^{z}S_{j+1}^{z}\right) ,
\end{equation}%
where $S_{j}^{x},\>S_{j}^{y}$ and $S_{j}^{z}$ are spin-1/2 operators at site
$j$. The model has a global SU(2) symmetry, and can be exactly solved by the
Bethe-ansatz method \cite{Bethe}. From the Bethe-ansatz solution, the energy
spectra of the system, its ground state properties, and the thermodynamics
of the system have been studied explicitly. Moreover, for the Heisenberg
model, the function $\langle S_{j}S_{j+r}\rangle $ now can be calculated by
using the generation function method \cite{JSato}. However, it is not our
motivation to readdress all these interesting issue. Instead, to illustrate
our scheme, let us forget all of the known results at the beginning.

Now suppose we do not know any information of long-range correlation
existing in the ground state of the system, we can only calculate the ground
state via numerical methods such as exact diagonalization and density-matrix
renormalization group\cite{DMRG}. With this state, we find that the reduced
density matrix of a single site is%
\begin{equation}
{\rho }_{i}=\left(
\begin{array}{cc}
1/2 & 0 \\
0 & 1/2%
\end{array}%
\right)
\end{equation}
so the entropy $S({\rho }_{i})=1$. For two separated spins, the reduced
density matrix has the form%
\begin{equation}
{\rho }_{i\cup j}=\left(
\begin{array}{llll}
u & 0 & 0 & 0 \\
0 & w & z & 0 \\
0 & z & w & 0 \\
0 & 0 & 0 & u%
\end{array}%
\right) ,  \label{eq:tsitematrix}
\end{equation}%
with $2u+2w=1$, $w\neq u$, and $z\neq 0$. For instance, if $|j-i|=5$, we
have $u\simeq 0.219$, $w\simeq 0.281$, and $z\simeq 0.062$\cite{Takahashi},
then $S({\rho }_{i\cup j}{)\simeq 1.969}$. The mutual information becomes
0.031. Differ from the usual long-range order for which the mutual
information tends to a constant as the distance between two blocks tends to
infinite, we find the mutual information here decays algebraically and
becomes zero finally. However, such a correlation still has a divergent
correlation length, so it still differs also from the disordered state for
which the correlation length is finite. To distinguish the correlation from
disorder and long-range order, we call such a correlation as long-range
correlation instead of long-range order.

Since $w\neq u$, according to our scheme, we can find that, as discussed in
the subsection III(A), the operator with diagonal long-range correlation is
just $\sigma _{i}^{z}=a_{i1}^{\dagger }a_{i1}-a_{i2}^{\dagger }a_{i2}$.

However, from the expression (\ref{eq:tsitematrix}), we notice the
off-diagonal element $z$ is nonzero too. So we predict that there exist some
kinds of off-diagonal long-range correlation. The potential operator can be
defined as%
\begin{eqnarray}
O_{i} &=&wa_{i1}^{\dagger }a_{i2}+w^{\ast }a_{i2}^{\dagger }a_{i1}  \nonumber
\\
&=&x\left( a_{i1}^{\dagger }a_{i2}+a_{i2}^{\dagger }a_{i1}\right) +iy\left(
a_{i1}^{\dagger }a_{i2}-a_{i2}^{\dagger }a_{i1}\right)  \nonumber \\
&=&xO_{i}^{x}-yO_{i}^{y}.  \label{eq:offdorder}
\end{eqnarray}%
Due to the exchange symmetry between $i$ and $j$, one can prove that $%
\left\langle O_{i}^{x}O_{i}^{y}\right\rangle =0$. From the expression (\ref%
{eq:tsitematrix}), we have
\begin{eqnarray}
\left\langle a_{i1}^{\dagger }a_{i2}a_{j1}^{\dagger }a_{j2}\right\rangle
&=&0, \\
\left\langle a_{i1}^{\dagger }a_{i2}a_{j2}^{\dagger }a_{j1}\right\rangle
&=&z, \\
\left\langle a_{i2}^{\dagger }a_{i1}a_{j1}^{\dagger }a_{j2}\right\rangle
&=&z, \\
\left\langle a_{i2}^{\dagger }a_{i1}a_{j2}^{\dagger }a_{j1}\right\rangle
&=&0.
\end{eqnarray}%
These equalities hold true if $\left\langle O_{i}^{x}O_{j}^{x}\right\rangle
=\left\langle O_{i}^{y}O_{j}^{y}\right\rangle $. Therefore, both $O_{i}^{x}$
and $O_{i}^{y}$ (or their linear combination) can be regarded as the
operator which has off-diagonal long-range correlation. From Eq. (\ref%
{eq:offdorder}), we can see that $O_{i}^{x}$ and $O_{i}^{y}$ are the
Jordan--Schwinger representation of the $x$ and $y$ components of Pauli
matrices $\sigma _{i}^{x}$ and $\sigma _{i}^{y}$, respectively. A careful
scrutiny may find that $u=w-z$, this equality which can be verified
numerically means that $\left\langle \sigma _{i}^{x}\sigma
_{j}^{x}\right\rangle =\left\langle \sigma _{i}^{z}\sigma
_{j}^{z}\right\rangle $.

Having obtained the operators for diagonal and off-diagonal long-range
correlation, we can find the mode of these operators by calculating the
corresponding correlation functions. Take the $\left\langle \sigma
_{i}^{z}\sigma _{i+r}^{z}\right\rangle $ as the example, we have $%
\left\langle \sigma _{i}^{z}\sigma _{i+1}^{z}\right\rangle =-0.59084$, $%
\left\langle \sigma _{i}^{z}\sigma _{i+2}^{z}\right\rangle =0.242716$, $%
\left\langle \sigma _{i}^{z}\sigma _{i+3}^{z}\right\rangle =-0.200992$, $%
\left\langle \sigma _{i}^{z}\sigma _{i+4}^{z}\right\rangle =0.138610$ \cite%
{Takahashi}., therefore the mode of the correlation function $\left\langle
\sigma _{i}^{z}\sigma _{i+r}^{z}\right\rangle $ is $\pi $. In the same way,
the mode of the operators $\sigma _{i}^{x}$ \ and $\sigma _{i}^{y}$ are $\pi
$ too.

Up to now, we have obtained the desired operators with antiferromagnetic
long-range correlations. This kind of correlation is consistent with the
definition of the Hamiltonian. The coefficient of each interaction term in
the Hamiltonian is positive. The three terms $S_{j}^{x}S_{j+1}^{x}$,$\
S_{j}^{y}S_{j+1}^{y}$, $S_{j}^{z}S_{j+1}^{z}$ favor to form
antiferromagnetic order in the ground state. From our analysis, we have $%
\left\langle \sigma _{i}^{x}\sigma _{j}^{x}\right\rangle =\left\langle
\sigma _{i}^{y}\sigma _{j}^{y}\right\rangle =\left\langle \sigma
_{i}^{z}\sigma _{j}^{z}\right\rangle $. The result can be explained from the
global SU(2) symmetry of the Hamiltonian.

\section{Summary}

\label{sec:sum} In summary, we have shown that the non-vanishing behavior of
the mutual information at a long-range distance means the existence of
long-range order. While the mutual information is operator-independent, we
could still find the potential diagonal or off-diagonal order parameter from
its spectra. A possible scheme to construct the order parameter was
provided. Our scheme, as shown by also three simple examples, can not only
find the diagonal order parameter, but also the off-diagonal order
parameter. To find a correct order parameter is very important in various
studies in the condensed matter physics, our scheme is therefore instructive
for many condensed matter theorists to explore new physics in unknown
quantum many-body systems.

\section{Acknowledgement}

SJ Gu thanks Jiu-Shu Shao and Dao-Xin Yao for helpful discussions on the
issue. This work is supported by the Earmarked Grant Research from the
Research Grants Council of HKSAR, China (Project No. HKUST3/CRF/09).

\end{document}